\documentclass[showpacs,prl,preprint]{revtex4}
\usepackage{amssymb}
\usepackage{graphicx}

\begin{document}

\title{Electrical Control of Exchange Bias Mediated by Graphene}
\author{Y. G. Semenov, J. M. Zavada, and K. W. Kim}
\address{Department of Electrical and Computer Engineering, North
Carolina State University, Raleigh, NC 27695-7911}

\begin{abstract}
The role of graphene in mediating the exchange interaction is
theoretically investigated when it is placed between two
ferromagnetic dielectric materials.  The calculation based on a
tight-binding model illustrates that the magnetic interactions at
the interfaces affect not only the graphene band structure but
also the thermodynamic potential of the system.  This induces an
indirect exchange interaction between the magnetic layers that can
be considered in term of an effective exchange bias field. The
analysis clearly indicates a strong dependence of the effective
exchange bias on the properties of the mediating layer, revealing
an effective mechanism of electrical control even at room
temperature.  This dependence also results in qualitatively
different characteristics for the cases involving mono- and
bilayer graphene.

\end{abstract}


\maketitle

The exchange bias is a phenomenon associated with the
unidirectional anisotropy of the exchange interaction at the
interface of ferromagnetic (FM) and antiferromagnetic (AFM)
materials \cite{Nogues99}. Widely used in the magnetic sensors,
magnetic random access memory, read heads of hard disk drives,
etc., its most important characteristic is the exchange bias field
that shifts the centrum of the hysteresis loop. Observation of
this phenomenon has generally been attributed to the use of
compounds whose FM component has a Curie temperature higher than
the N\'{e}el temperature of the AFM counterpart. Hence the
structure, when cooled in the presence of an external magnetic
field, first aligns the FM layer which in turn orders the AFM
component accordingly. In the proximity of an AFM pinning layer at
or below the N\'{e}el temperature, the FM material thus
experiences the unidirectional anisotropy induced by the exchange
interaction. The temperature dependence can also lead to the
unusual effect of spontaneous magnetization reversal through an
interplay between the external magnetic field and the exchange
bias field aligned in the antiparallel direction \cite{Li06}. In
some cases, the exchange bias field may degrade gradually after
multiple recycling of the system through the consequential
hysteresis loops (i.e., the so-called training effect in the
polycrystalline magnetic materials)
\cite{Nogues99,Zhang01,Hochstrat02}.

As all the properties described above are related to the magnetic
phenomena or the magnetic field repercussion, it raises a natural
question: Can an electric field control the exchange bias field at
the magnetic interfaces? One potential solution may be to utilize
the multiferroic films in place of the FM layers as it has been
discussed recently in Ref.~\cite{Bea08}. However, the variety of
multiferroic materials that can operate at room temperature have
so far been limited \cite{Bea08,Duan06}. On the other hand, a
completely different approach may be possible by exploiting the
unique properties of atomically thin graphite (i.e., graphene)
when placed at the interface between two magnetic dielectric
layers. Previous analyses demonstrated that the novel phenomena
induced by the exchange interaction between FM layers and
monolayer graphene (MLG) \cite{SZK07,Brataas08} or bilayer
graphene (BLG) \cite{SZK08} depend sensitively on the detailed
electronic features. It certainly offers a ground to explore
electrical control of the magnetic interactions, for which the
graphene electrons play a vital role.

In this work, we explore the exchange interaction of dielectric
magnetic layers mediated by graphene and its manipulation by
electrical control. Our theoretical result clearly illustrates
that this interaction can be expressed in terms of the effective
exchange bias field. It also reveals that the strength of the
effective bias field depends critically on the electronic
properties of graphene, particularly the position of the
electro-chemical potential $\mu $ (i.e., the carrier density),
amenable for electrical control \cite{Geim07}.  Numerical
estimates of the predicted phenomenon are provided for an order of
magnitude analysis, illustrating its potential significance in
spintronic applications.

For accurate understanding, it is useful to identify the
underlying differences between those involving MLG and BLG. In the
case of MLG, the electron spins polarized by one FM layer interact
with the other FM layer mediating their parallel alignment.
Therewith the indirect interaction between FM layers amplifies
with an increase in the electron density in MLG. On the other
hand, the two carbon layers in BLG interact primarily with the
nearest magnetic dielectric, respectively, leading to the
possibility of inter-layer symmetry break through nonidentical
exchange interactions \cite{SZK08}. For example, the difference in
the interaction can be achieved by varying the mutual orientation
of FM layers that are in contact with BLG at the top and bottom
interfaces. In addition to the modified BLG band structure as
discussed in Ref.~\cite{SZK08} (similar to the case of an external
electrical
bias~\cite{Novoselov06,Fal'ko06,Castro06,McDonald07,Cann07}), the
impact of this asymmetry is more fundamental by energetically
favoring an antiparallel configuration in the magnetization
orientation of two magnetic layers \cite{Comm}. Hence, both MLG
and BLG can be considered as mediating the exchange interaction
(and, thus, the exchange bias) between FM layers, each with a
different preference for the bias field orientation.
%As the properties of
%graphene can be modulated electrically (for example, the location
%of electro-chemical potential $\mu $, etc.), a corresponding
%opportunity is clearly open to the subsequent indirect interaction
%and the exchange bias.

We begin with a brief analysis of the graphene electronic band
structure sandwiched between two magnetic dielectrics. The
magnetization belonging to each dielectrics, or more precisely to
the proximate strata of FM (or AFM) films in contact with graphene
\cite{Meiklejohn57}, is defined as $\mathbf{M}_{1}$ and
$\mathbf{M}_{2}$, respectively. A particularly interesting feature
appears for ferromagnetic dielectric layers (FDLs) with an
identical magnitude of the magnetic moments ($\left\vert
\mathbf{M}_{1}\right\vert =\left\vert \mathbf{M}_{2}\right\vert
\equiv M$) so that the total magnetization can be varied from zero
to $2M$. If the absence of electrically induced asymmetry is
further assumed, then only the relative misalignment of
$\mathbf{M}_{1}$ and $\mathbf{M}_{2}$ provides the cause of
symmetry break between two graphene layers in the case of BLG.

Figure~1 schematically illustrates the specific structure under
consideration. It resembles the ferromagnet-metal-ferromagnet
hybrid structures  that reveal a giant magnetoresistance owing to
the spin-dependent in-plane conductivity \cite{Fert88}. In our
case, graphene substitutes the metallic film. The magnetization
$\mathbf{M}_{1}$ of the bottom FDL is assumed to pinned along the
$x$ direction by an AFM substrate with a sufficiently strong
exchange bias field. The top FDL is constructed from the same
material but its magnetization vector $\mathbf{M}_{2}$ can be
rotated in the $x$-$y$ plane (by an external or "internal"
magnetic field) forming an angle $\theta $ with $\mathbf{M}_{1}$.

The influence of the FDL magnetization on the graphene electronic
structure can be realized in actual structures through either the
exchange interaction with magnetic ions (assuming an overlap
between the carbon $\pi $-orbitals and unfilled shells of the
magnetic ions in the FDLs) or an interaction via the ligands of
FDLs. Thus, the problem can be modeled in the mean field
approximation \cite{SZK07,Brataas08} with the Hamiltonian
\begin{equation}
\mathcal{H}^{(n)}=\mathcal{H}_{G}^{(n)}+\mathcal{H}_{ex}^{(n)}\,,  \label{f1}
\end{equation}%
where $\mathcal{H}_{G}^{(n)}$ is the spin-independent MLG ($n=1$) or BLG ($%
n=2$) Hamiltonian. The remaining term of Eq.~(\ref{f1}) describes
the energy of an electron spin $\mathbf{S}$ in the effective
fields (in units of energy) $\alpha \mathbf{M}_{1}$ and $\alpha
\mathbf{M}_{2}$ of the proximate FDLs, where parameter $\alpha $
is proportional to the carrier-ion exchange constants as evaluated
in Refs.~\cite{SZK07} and \cite{Brataas08}. Hence, it can be
written for MLG as
\begin{equation}
\mathcal{H}_{ex}^{(1)}=\alpha (\mathbf{M}_{1}+\mathbf{M}_{2})\mathbf{S} \,,
\label{f1a}
\end{equation}%
while the corresponding expression for BLG is
\begin{equation}
\mathcal{H}_{ex}^{(2)}=\mathcal{P}_{1}\alpha \mathbf{M}_{1}\mathbf{S+}%
\mathcal{P}_{2}\alpha \mathbf{M}_{2}\mathbf{S} \,.  \label{f1b}
\end{equation}%
In Eq.~(\ref{f1b}), the projection operator $\mathcal{P}_{1}$ ($\mathcal{P}%
_{2}$) is 1 for the electron localized at the bottom (top) carbon monolayer
and 0 otherwise.

In the case of low energy electronic excitations, the
tight-binding approximation accurately describes the band spectra
of graphene in the vicinities of each valley $K$ and $K^{\prime }$
\cite{Slonczewski58,Fal'ko06}. Hence, we adopt the tight-binding
Hamiltonian near the valley extrema augmented to account for the
electron exchange energy in the form of Eq.~(\ref{f1a}) or
Eq.~(\ref{f1b}). The qualitative difference between the two spin
Hamiltonians [i.e., Eqs.~(\ref{f1a}) and (\ref{f1b})] implies that
$\mathcal{H}_{ex}^{(1)}=2\alpha M\cos (\theta /2)S_{M}$ ($ S_{M}$
is the electron spin projection on the direction of sum
$\mathbf{M}_{1}+\mathbf{M}_{2}$) commutes with
$\mathcal{H}_{G}^{(1)}$ while this is not the case for
$\mathcal{H}_{ex}^{(2)}$ and $\mathcal{H}_{G}^{(2)}$. As a result,
MLG remains gapless with four non-degenerate branches $\varepsilon
_{b,\sigma }^{(1)}(\mathbf{k})$ near the $K$ and $K^{\prime }$
points (except the case of $\theta =\pi $ that corresponds to
$\mathcal{H}_{ex}^{(1)}=0$). They are identical for the conduction
($b=+1$) and valence ($b=-1$) bands (i.e., the reflection symmetry
about $\varepsilon =0$) and isotropic with respect to the valley
centrum $k=0$. The degeneracy of spin doublet is lifted as stated,
which leads to the expression
\begin{equation}
\varepsilon _{b,\sigma }^{(1)}(\mathbf{k})=b \hbar v_{F}k+\sigma
\alpha M\cos \frac{\theta }{2}  \,. \label{e1}
\end{equation}%
Here, $v_{F}=10^{8}$ cm/s is the Fermi velocity in MLG, the
subband index $\sigma = \pm 1$, and $\theta$ is the angle between
$\mathbf{M}_{1}$ and $ \mathbf{M}_{2}$ ($0 \leq \theta \leq \pi
$).  The impact of the spin Hamiltonian on the MLG band structure
is clearly illustrated in Figs.~2(a) and 2(b) for two opposite
cases of $\mathbf{M}_{1}$/$ \mathbf{M}_{2}$ alignment.  An
important point to note is that, while MLG remains gapless
independent of $\theta$ and $\alpha M$ following the qualitative
discussion given above, the contact (i.e., zero gap) points are
shifted to the circle of radius $k_{c}=(|\alpha| M/\hbar
v_{F})\cos \theta /2$ in the $k_x$-$k_y$ plane (measured from the
centrum of the $K$ or $K^{\prime }$ point).   Once $\theta = \pi$
and $ \mathbf{M}_{1}+\mathbf{M}_{2}$ becomes zero in
Eq.~(\ref{f1a}), the band structure returns to that of unaltered
MLG as shown in Fig.~2(b). Consequently, the density of states for
MLG interacting with FM layers modifies to
\begin{equation}
\rho ^{(1)}(\varepsilon )= 2 \frac{\max \{\left\vert \varepsilon
\right\vert ,\left\vert \alpha \right\vert M\cos \frac{\theta
}{2}\}}{\pi \hbar ^{2}v_{F}^{2}} \,,  \label{e2}
\end{equation}%
that is non-zero even at $\varepsilon=0$.

For BLG with the Hamiltonian $ \mathcal{H}^{(2)}$, a qualitatively
different situation is realized due to the interlayer electron
transitions with the corresponding matrix element $\gamma
_{1}=0.4$~eV. Following the approach discussed in
Ref.~\cite{SZK08}, the BLG energy spectra can be obtained in terms
of eight energy branches $\varepsilon _{b,\lambda ,\sigma
}^{(2)}(\mathbf{k})$ for each valley. Along with $b$ and $\sigma$
described above, an additional index $\lambda = \pm 1$ is
introduced to distinguish four low-energy bands $\varepsilon
_{b,-1,\sigma }^{(2)}(\mathbf{k})$ from the other four excited
states with energies $\left\vert \varepsilon _{b,1,\sigma
}^{(2)}(\mathbf{k})\right\vert \gtrsim \gamma _{1}$. As it is
convenient to normalize the parameters in units of $ \gamma _{1}$,
the dimensionless momentum $\mathbf{p}\equiv \hbar v_{F}
\mathbf{k}/\gamma _{1}$ and the effective field $\mathbf{G}\equiv
\alpha \mathbf{M}/\gamma _{1}$ are used hereinafter. Then, the
energy bands can be expressed as
\begin{equation}
\varepsilon _{b,\lambda ,\sigma }^{(2)}(\mathbf{p})=b\gamma
_{1}\sqrt{p^{2}+ \frac{G^{2}}{4}+\frac{1}{2}\left( 1 + \sigma
G\cos \frac{\theta }{2}+\lambda W_{\sigma }(p)\right) } \,,
\label{f3}
\end{equation}%
where
\begin{equation}
W_{\sigma }(p)=\sqrt{\left( 1 + \sigma G\cos \frac{\theta
}{2}\right) ^{2}(1+4p^{2})+\left( 2pG\sin \frac{\theta }{2}\right)
^{2}} \,. \label{f4}
\end{equation}%
Four solutions with $\varepsilon _{+1,\lambda ,\sigma }>0$
correspond to the conduction bands, while their mirror images with
respect to the zero energy describe the highest valence bands
($b=-1$).

In contrast to MLG, the calculation clearly illustrates the
presence of an energy gap $\varepsilon_{g}$ between the lowest
conduction band and the highest valence band as soon as $\theta
\neq 0$. When the orientation of $\mathbf{M}_{1}$ and
$\mathbf{M}_{2}$ is in a parallel alignment ($\theta =0$), the net
effect of the exchange interaction simply lifts the two-fold spin
degeneracy resulting in two pairs of spin-split bands that cross
each other at $p=\sqrt{G(1+G/2)/2}$ as shown in Fig.~2(c).
However, once they are misaligned, the electronic bands become of
mixed spin character (e.g., with both parallel and antiparallel
components to the $x$ direction). Subsequent anti-crossing opens
up the gap that progressively grows with $\theta $. At $ \theta
=\pi $ in Fig.~2(d) (i.e., $\mathbf{M}_{1}=-\mathbf{M}_{2}$), the
gap reaches the maximum while the energy bands regain the two-fold
degeneracy similar to MLG \cite{SZK08}. Accordingly, the BLG
density of states $\rho ^{(2)}(\varepsilon )$ is zero for $
\left\vert \varepsilon \right\vert \leq \varepsilon_{g}/2$, where
$\varepsilon_{g}=\gamma_1 G\sin \frac{ \theta
}{2}/\sqrt{1+G^{2}+2G\cos \frac{\theta }{2}}$.  As the density of
states determines the characteristic carrier occupancy, the
calculation result strongly indicates that the total electronic
energy of the structure can be controlled by the orientation of
$\mathbf{M}_{1}$ and $ \mathbf{M}_{2}$ or, more specifically, by
the angle $\theta$. This point is schematically illustrated in
Fig.~2 through the comparison of shaded regions representing the
occupied states when the electro-chemical potential $\mu$ (dashed
line) is hypothetically located at $\varepsilon =0$.

Now we can take into account the entropy effects at finite
temperature along with the band modification [Eqs.~(\ref{e1}) and
(\ref{f3})] by evaluating the thermodynamic potential
\begin{equation}
\Omega _{n}(\theta )=- 2 k_{B}T\sum\limits_{\{m\}}\sum\limits_{\mathbf{\ k}%
} \ln \left[ 1+\exp \left( \frac{\mu -\varepsilon _{\{m\}}^{(n)}(\mathbf{k})%
}{k_{B}T}\right) \right] \,,  \label{f7}
\end{equation}%
where $k_{B}$ is the Boltzmann constant, $T$ is the temperature,
$n=1,2$ denotes the cases of MLG and BLG, respectively, and
$\{m\}$ collectively represents the band indices described
earlier. Additionally, the factor of 2 comes from the summation
over the valleys $K$ and $K^{\prime }$.  One remarkable outcome
from the calculations of thermodynamic potential [Eq.~(\ref{f7})]
is the universal scalability of the dependence on angle $\theta $
for both MLG and BLG:
\begin{equation}
\frac{\Delta \Omega _{n}(\theta )}{\Delta \Omega _{n}(\pi )}=\frac{1}{2}%
\left( 1-\cos \theta \right) ,  \label{f8}
\end{equation}%
where $\Delta \Omega _{n}(\theta )=\Omega _{n}(\theta )-\Omega
_{n}(0)$. In other words, the relative change in the thermodynamic
potential [Eq.~(\ref{f8})] is independent of all parameters except
$ \cos \theta$ in spite of its seemingly complex expression [see
Eqs.~(\ref{e1}), (\ref{f3}), and (\ref{f7})].

Clearly, Eq.~(\ref{f8}) indicates that the free energy of the
system either increases or decreases as the magnetization of the
free FDL ($\mathbf{M}_{2}$) rotates from the parallel ($\theta
=0$) to the antiparallel ($\theta = \pi$).  Since a lower free
energy is favored, it means that $\mathbf{M}_{2}$ is inclined to
take one of the orientations depending on the sign of $\Delta
\Omega _{n}(\pi )$.  This preference on the magnetization
orientation can also be described in terms of the magnetic energy
of the FM layer when it is subject to a magnetic field: i.e.,
$\Omega _{M}(\theta )=-\mathbf{H}_{eb} \mathcal{ M}_{2} =
-H_{eb}\mathcal{M}_{2}\cos \theta $, where the $x$ axis is chosen
as the reference direction for the magnetic field and
$\mathcal{M}_{2}=M_{2}A_{0}t_{F}$ is the total magnetic moment of
the top FDL with $A_{0}$ and $t_{F}$ its area and thickness.
Through comparison with Eq.~(\ref{f8}), one can readily deduce the
strength of the effective magnetic field $H_{eb}^{(n)}$ that is
essentially the {\em exchange bias field mediated by graphene},
\begin{equation}
H_{eb}^{(n)}=\frac{\Delta \Omega _{n}(\pi )}{2\mathcal{M}_{2}} \,.
\label{f9}
\end{equation}%
As in the case of conventional exchange bias, the strength of
$H_{eb}^{(n)}$ is inversely proportional to the thickness $t_{F}$
\cite{Nogues99}.  However, one distinguishing feature of the
exchange bias field given in Eq.~(\ref{f9}) is its dependence on
the electronic properties of the graphene layer, particularly the
position of the electro-chemical potential $\mu $ that can be
readily modulated by the gate bias ($V_{g1} $, $V_{g2}$; see
Fig.~1). This also leads to qualitatively different
characteristics for MLG and BLG that the calculation of $\Delta
\Omega _{n}(\pi )$ highlights.

Firstly, the signs of $H_{eb}^{(1)}$ and $H_{eb}^{(2)}$ are
different at least in the range $\left\vert \mu \right\vert \leq
0.3 \gamma_1$.  While MLG tends to establish $\mathbf{M}_{2}$
parallel to $\mathbf{M}_{1}$ ($H_{eb}^{(1)} > 0$), BLG favors the
antiparallel alignment ($H_{eb}^{(2)} < 0$) as pointed out in our
discussion earlier. Secondly, a shift of $\mu $ from the graphene
charge neutrality point ($\varepsilon = 0 $) affects the strengths
of the exchange bias fields in the opposite directions. Namely,
the magnitude of $ H_{eb}^{(1)}$ gradually increases with $ | \mu
|$, whereas that of $H_{eb}^{(2)}$ is at the maximum at $ \mu =0$
and decreases to zero.  A similar pattern is also observed in the
temperature dependence. As $T$ goes up, $ H_{eb}^{(1)}$ becomes
stronger and $H_{eb}^{(2)}$ weaker when $\mu$ is at or around the
graphene charge neutrality point. The properties of $H_{eb}^{(1)}$
can be readily explained by considering the concentration
variation of the conduction electrons and valence holes that can
be spin polarized and therefore establish an effective field
affecting equally the bottom and top FM layers. Hence, the
parallel alignment of two ferromagnets are energetically favored
and the magnitude of the indirect exchange bias field increases
with the mediating carrier density (i.e., $| \mu |$) when MLG is
used. On the other hand, the effect of $H_{eb}^{(2)}$ can be
thought in terms of the band structure modification such as the
size of the gap. As a larger gap pushes the energies of the
occupied states lower, the thermodynamic potential decreases
likewise.  This prefers a larger angle $\theta$ with $H_{eb}^{(2)}
< 0$ (i.e., antiparallel). Increasing $\left\vert \mu \right\vert
$ and $T$ diminishes the effect and, thus, the magnitude of
$H_{eb}^{(2)}$.

The aforementioned characteristics can be captured by an
approximate expression in the limit of small $G$,
\begin{equation}
\Delta \Omega _{n}(\pi ) \simeq NG^{2}f_{n}(\mu ,T),  \label{f10}
\end{equation}%
where $N$ is the number of graphene primitive cells at the
interface and the factor $f_{n}(\mu ,T)$ provides the specific
dependence on $\mu$ and $T$ for MLG ($n=1$) and BLG ($n=2$).  The
dependence of the exchange bias field [Eqs.~(\ref{f9}) and
(\ref{f10})] on the square of the effective field $G$ is not
surprising as the indirect interaction between the bottom
($\mathbf{M}_{1}$) and the top ($\mathbf{M}_{2}$) ferromagnets
involves two interfaces with graphene. Figure~3 shows $ f_{n}(\mu
,T) $ vs.\ $\left\vert \mu \right\vert $ evaluated at three
different temperatures. As expected, the temperature factor has a
considerable influence around $\mu =0$ but its role diminishes
significantly for $ \left\vert \mu \right\vert
> 0.2 \gamma _{1}$ ($\gamma_1 = 0.4$~eV).  More crucial is
the possibility of controlling $ f_{n}(\mu ,T)$ (thus,
$H_{eb}^{(n)}$) over a wide range even at room temperature. As
evident from the figure, the shift of $\mu \simeq \pm 0.15\gamma
_{1}$ can change the strength of $H_{eb}^{(n)}$ by about a factor
of two for both cases mediated by MLG and BLG. Based on these
results, a rough estimate can be made for the magnitude of
$H_{eb}^{(n)}$. Assuming $\left\vert f_{n}\right\vert \approx
1$~meV (a typical number from Fig.~3) \cite{Katayama06}, a high
temperature FM dielectric such as yttrium iron garnet, and $G=0.1$
for the graphene interface ($\gamma_1 G = 40$~meV) \cite{SZK08},
one can find $t_{FM} \times H_{eb}^{(n)}\approx 1100$~Oe, where
$t_{FM}$ is the FM layer thickness in nm (e.g.,
$H_{eb}^{(n)}\approx 1100$~Oe at $t_{FM} =1$~nm).  Electrical
control of the exchange bias in the estimated range can be of
practical importance in a number of spintronic applications.

In summary, our theoretical analysis shows that graphene can
mediate the indirect interaction of magnetic layers resulting in
an effective exchange bias.  Through the dependence on the
graphene electro-chemical potential, it is also clearly identified
that the effective exchange bias can be modulated electrically
over a wide range even at room temperature. The numerical
estimation indicates the potential significance of the proposed
phenomenon in practical applications.

This work was supported in part by the US Army Research Office and the FCRP
Center on Functional Engineered Nano Architectonics (FENA).

\newpage

\begin{center}
\begin{figure}[tbp]
\includegraphics[scale=.55,angle=0]{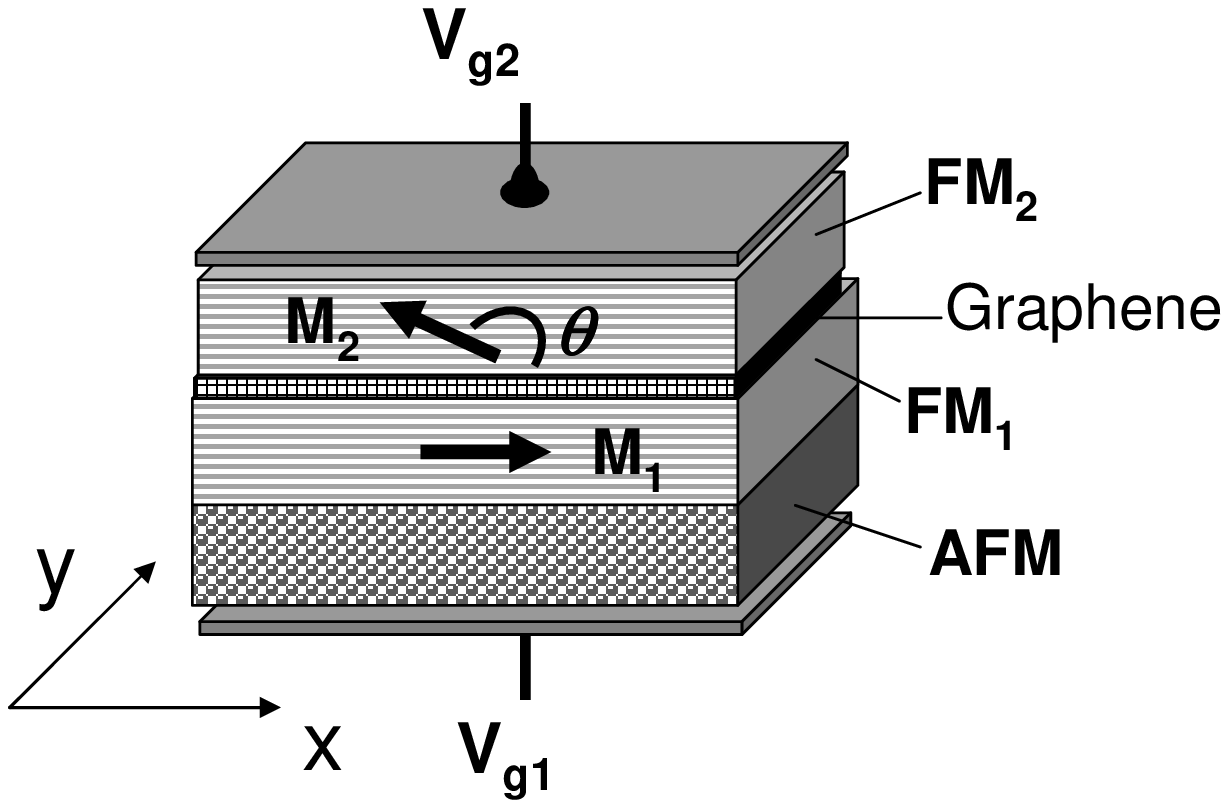}
\caption{Schematic illustration of graphene sandwiched between two
FM dielectric layers $FM_{1}$ and $FM_{2}$ of magnetization
$\mathbf{M}_{1}$ and $\mathbf{M}_{2}$.  While $\mathbf{M}_{1}$ is
pinned by an AFM layer along the $x$ direction, $\mathbf{M}_{2}$
can rotate on the $x$-$y$ plane with $ \protect\theta$ specifying
the angle between them. The structure can be placed between
metallic contacts providing electrical control with the gate
voltages $V_{g1}$ and $V_{g2}$.}
\end{figure}
\end{center}

\newpage

\begin{center}
\begin{figure}[tbp]
\includegraphics[scale=0.8,angle=0]{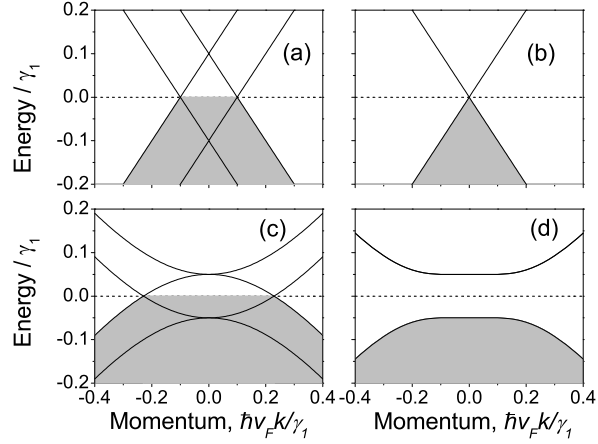}
\caption{Energy band diagram of (a,b) mono- and (c,d) bilayer
graphene when sandwiched between two FM materials as shown in
Fig.~1.  For (a) and (c), the angle $\theta$ between
$\mathbf{M}_{1}$ and $\mathbf{M}_{2}$ is zero (i.e., parallel
alignment) and no gap exists between the bands.  In the cases of
(b) and (d), $\theta = \pi$ (antiparallel) and the bands are
doubly degenerate.  The shaded regions schematically represent the
occupied valence bands with $\protect\mu =0$ (dashed line) at $T
=0$. The effective field $G = 0.1$ is assumed for the calculation
and the momentum is measured from the $K$ or $K^{\prime }$ valley
centrum.  $\gamma_1 = 0.4$~eV and $v_F = 10^8$~cm/s.}
\end{figure}
\end{center}

\newpage

\begin{center}
\begin{figure}[tbp]
\includegraphics[scale=.8,angle=0]{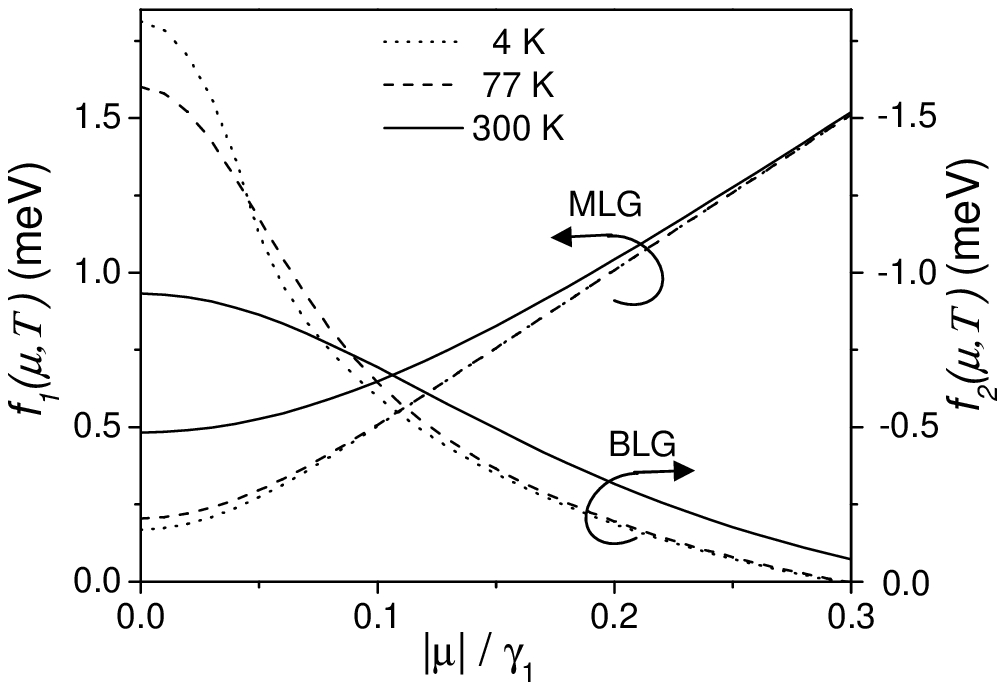}
\caption{Factor $ f_{n}(\protect\mu ,T)$ [$ \simeq \Delta \Omega
_{n}(\protect\pi) /NG^{2}$] vs.\ $\left\vert \mu \right\vert $
evaluated at three different temperatures ($n=1$ for MLG; $n=2$
for BLG).  This factor essentially determines the dependence of
exchange bias field on $ \mu$ and $T$. }
\end{figure}
\end{center}


\clearpage

\begin{thebibliography}{99}
\bibitem{Nogues99} J. Nogu\'{e}s and I. K. Schuller, J. Magn. Magn. Mater.
\textbf{192}, 203 (1999).

\bibitem{Li06} Z. P. Li \textit{et al.}, Phys. Rev. Lett. \textbf{96}, 137201 (2006).

%\bibitem{Li06} Z. P. Li, J. Eisenmenger, C. W. Miller, and I. K. Schuller,
%Phys. Rev. Lett. \textbf{96}, 137201 (2006).

\bibitem{Zhang01} K. Zhang, T. Zhao, and M. Fujiwara, J. Appl. Phys. \textbf{%
89}, 6010 (2001).

\bibitem{Hochstrat02} A. Hochstrat, Ch. Binek, and W. Kleemann, Phys. Rev. B
\textbf{66}, 092409 (2002).

\bibitem{Bea08} H. B\'{e}a \textit{et al.}, Phys. Rev. Lett. \textbf{100}, 017204 (2008).

%\bibitem{Bea08} H. B\'{e}a, M. Bibes, F. Ott, B. Dup\'{e}, X.-H. Zhu, S.
%Petit, S. Fusil, C. Deranlot, K. Bouzehouane, and A. Bart\'{e}l\'{e}my,
%Phys. Rev. Lett. \textbf{100}, 017204 (2008).

\bibitem{Duan06} C.-G. Duan \textit{et al.}, Phys. Rev. Lett. \textbf{97}, 047201 (2006).

\bibitem{SZK07} Y. G. Semenov, K. W. Kim, and J. Zavada, Appl. Phys. Lett.
\textbf{91}, 153105 (2007).

\bibitem{Brataas08} H. Haugen, D. Huertas-Hernando, and A. Brataas, Phys.
Rev. B \textbf{77}, 115406 (2008).

\bibitem{SZK08} Y. G. Semenov, J. M. Zavada, and K. W. Kim, Phys. Rev. B
\textbf{77}, 235415 (2008).

\bibitem{Geim07} A. K. Geim and K. S. Novoselov, Nat. Mater. \textbf{6}, 183
(2007).

\bibitem{Novoselov06} K. S. Novoselov \textit{et al.}, Nat. Phys. \textbf{2}, 177 (2006).

%\bibitem{Novoselov06} K. S. Novoselov, E. McCann, S. V. Morozov, V. I.
%Fal'ko, M. I. Katsnelson, U. Zeitler, D. Jiang, F. Schedin, and A. K. Geim,
%Nat. Phys. \textbf{2}, 177 (2006).

\bibitem{Fal'ko06} E. McCann and V. I. Fal'ko, Phys. Rev. Lett. \textbf{96},
086805 (2006).

\bibitem{Castro06} E. V. Castro \textit{et al.}, Phys. Rev. Lett. \textbf{99}, 216802
(2007).

%\bibitem{Castro06} E. V. Castro, K. S. Novoselov, S. V. Morozov, N. M. R.
%Peres, J. M. B. Lopes dos Santos, J. Nilsson, F. Guinea, A. K. Geim, and A.
%H. Castro Neto, Phys. Rev. Lett. \textbf{99}, 216802 (2007).

\bibitem{McDonald07} H. Min \textit{et al.}, Phys. Rev. B \textbf{75}, 155115 (2007).

%\bibitem{McDonald07} H. Min, B. Sahu, S. K. Banerjee, and A. H. MacDonald,
%Phys. Rev. B \textbf{75}, 155115 (2007).

\bibitem{Cann07} E. McCann, D. S. L. Abergel, and V. I. Fal'ko, Eur. Phys.
J. Spec. Top. \textbf{148}, 91 (2007).

\bibitem{Comm} Note that the prospective effect is in accordance with other
physical phenomena (such as the Jahn-Teller effect, Peierls transition,
etc.) of electronic stabilization through the spontaneously broken symmetry.

%\bibitem{Novoselov07} K. S. Novoselov, Y. Zhang, A. K. Geim, Y. Zhang, S. V.
%Morozov, H. L. Stormer, U. Zeittler, J. C. Maan, G. S. Boebinger,
%P. Kim, and A. K. Geim, Science \textbf{315}, 1379 (2007).

\bibitem{Meiklejohn57} W. H. Meiklejohn and C. P. Bean, Phys. Rev. \textbf{%
105}, 904 (1957).

\bibitem{Fert88} See, for example, M. N. Baibich \textit{et al.},
Phys. Rev. Lett. \textbf{61}, 2472 (1988).

%\bibitem{Fert88} See, for example, M. N. Baibich, J. M. Broto, A. Fert, F.
%N. V. Dau, F. Petroff, P. Eitenne, G. Creuzet, A. Frederich, and J.
%Chazelas, Phys. Rev. Lett. \textbf{61}, 2472 (1988).

\bibitem{Slonczewski58} J. C. Slonczewski and P. R. Weiss, Phys. Rev.
\textbf{109}, 272 (1958).

\bibitem{Katayama06} An interlayer exchange coupling of the same order of
magnitude was found for Fe-Fe through MgO.  See T. Katayama
\textit{et al.}, Appl. Phys. Lett. \textbf{89}, 112503 (2006).

%\bibitem{Melo97} L. V. Melo, L. M. Rodrigues, and P. P. Freitas, IEEE Trans.
%Magn. \textbf{33}, 3295 (1997).

%\bibitem{Matsuyama97} K. Matsuyama \textit{et al.}, IEEE Trans. Magn.
%\textbf{33}, 3283 (1997).

%\bibitem{Matsuyama97} K. Matsuyama, H. Asada, S. Ikeda, and K. Taniguchi,
%IEEE Trans. Magn. \textbf{33}, 3283 (1997).
\end{thebibliography}
\end{document}